\documentclass[a4paper]{article}
\usepackage{amsmath,amssymb,theorem,fullpage}
\usepackage[nosort]{cite}

\let\frac\undefined

\numberwithin{equation}{section}

\def\Maketitle{{\def\newpage{}\maketitle}}

\def\eq#1{\begin{equation}#1\end{equation}}

\def\Align#1{\begin{align}#1\end{align}}
\def\AlignStar#1{\begin{align*}#1\end{align*}}
\def\Aligned#1{\begin{aligned}#1\end{aligned}}

\def\Multline#1{\begin{multline}#1\end{multline}}
\def\Matrix#1{\begin{matrix}#1\end{matrix}}

\def\Cases#1{\begin{cases}#1\end{cases}}

\def\Tr{\mathop{\rm Tr}\nolimits}
\def\Re{\mathop{\rm Re}\nolimits}
\def\Im{\mathop{\rm Im}\nolimits}
\def\sign{\mathop{\rm sign}\nolimits}
\def\const{\mathop{\rm const}\nolimits}
\def\cF{{\cal F}}
\def\cH{{\cal H}}
\def\ve{\varepsilon}
\def\sh{\mathop{\rm sh}\nolimits}
\def\ch{\mathop{\rm ch}\nolimits}
\def\lcolon{\mathopen:}
\def\rcolon{\mathclose:}
\def\C{{\mathbb{C}}}
\def\R{{\mathbb{R}}}
\def\Z{{\mathbb{Z}}}
\def\P{{\cal P}}
\def\Q{{\cal Q}}
\def\llangle{\mathopen{\langle\!\langle}}
\def\rrangle{\mathclose{\rangle\!\rangle}}
\def\vR{{\check R}}
\def\vH{{\check H}}
\def\vG{{\check G}}

\newtheorem{theorem}{Theorem}

\def\RSOS#1{$\rm RSOS^{(#1)}$}

\def\W#1(#2,#3;#4,#5|#6){\mathop{W#1}\left[\Matrix{#5&#4\\#2&#3}\bigg|
                        \Matrix{\displaystyle{#6}}\right]}
\def\tW#1(#2,#3;#4,#5|#6){\mathop{\tilde W#1}\left[\Matrix{#5&#4\\#2&#3}\bigg|
                        \Matrix{\displaystyle{#6}}\right]}
\def\S#1(#2,#3;#4,#5|#6){\mathop{S#1}\left[\Matrix{#5&#4\\#2&#3}\bigg|
                        \Matrix{\displaystyle{#6}}\right]}

\makeatletter
\def\section{\@startsection{section}{1}{\z@}%
                                   {-3.5ex \@plus -1ex \@minus -.2ex}%
                                   {2.3ex \@plus.2ex}%
                                   {\normalfont\normalsize\bfseries}}
\def\subsection{\@startsection{subsection}{2}{\z@}%
                                     {-3.25ex\@plus -1ex \@minus -.2ex}%
                                     {1.5ex \@plus .2ex}%
                                     {\normalfont\normalsize\bfseries}}
\def\@seccntformat#1{\csname the#1\endcsname.~~}
\long\def\@makecaption#1#2{%
  \vskip\abovecaptionskip
  \sbox\@tempboxa{\small#1. #2}%
  \ifdim \wd\@tempboxa >0.9\hsize
  {\leftskip=0.05\hsize\rightskip=0.05\hsize\relax\small
    #1. #2\par}
  \else
    \global \@minipagefalse
    \hb@xt@\hsize{\hfil\box\@tempboxa\hfil}%
  \fi
  \vskip\belowcaptionskip}
\def\Appendix{\appendix
  \def\@seccntformat##1{Appendix~\csname the##1\endcsname.~~}}

\let\over\@@over
\let\atop\@@atop
\let\above\@@above
\let\overwithdelims\@@overwithdelims
\let\atopwithdelims\@@atopwithdelims
\let\abovewithdelims\@@abovewithdelims

\makeatother

\begin{document}

\def\i{{\rm i}}
\def\e{{\rm e}}

\title{Notes on scaling limits in SOS models,\\
local operators and form factors}

\author{Michael Lashkevich,\\[\medskipamount]
\it Landau Institute for Theoretical Physics,\\
\it 142432 Chernogolovka of Moscow Region, Russia}

\date{}

\rightline{hep-th/0510058}
\Maketitle

\begin{abstract}
Scaling limits of the SOS and RSOS models in the regime~III are
considered. These scaling limits are believed to be described by the
sine-Gordon model and the restricted sine-Gordon models (or perturbed
minimal conformal models) respectively. We study two different scaling
limits and establish the correspondence of the scaling local height
operators to exponential or primary fields in quantum field theory. An
integral representation for form factors is obtained in this way. In the
case of the sine-Gordon model this reproduces Lukyanov's well known
representation. The relation between vacuum expectation values of local
operators in the sine-Gordon model and perturbed minimal models is also
discussed.
\end{abstract}

%%%%%%%%%%%%%%%%%%%%%%%%%%%%%%%%%%%%%%%%%%%%%%%%%%%%%%%%%%%%%%%%%%%%%%%%
\section{Introduction}

The solid-on-solid (SOS) and restricted solid-on-solid (RSOS) models in
the vicinity of the critical points are well known to be related to the
perturbed conformal field theory~\cite{Andrews:1984af,FB,DJKMO,PN}. The scaling
behavior of the order parameters in the RSOS models was analyzed by
Huse~\cite{Huse}. It was shown that it is possible to define the order
parameters in such a way that they possess definite scaling dimensions
that coincide to those of appropriate conformal fields in the respective
models of the conformal field theory.

The consideration by Huse was based on the notion of the local height
probabilities, which are the simplest correlation functions in the SOS
and RSOS models. In the transfer matrix approach these quantities are
vacuum expectation values of local height operators. These can be
naturally generalized to the multi-point local height operators. The
expectation values and form factors (i.e.\ matrix elements in the basis
of eigenvalues of the transfer matrix) can be obtained by use of the
free field representation~\cite{LukPug,MW}.

Earlier~\cite{Lash97} it was shown on the example of the six-vertex
model that the free field representation makes it possible to obtain the
scaling limit of local operators. In this note we discuss the scaling
limit of the form factors of local height operators in the regime~III of
the SOS and RSOS models. We use another set of order parameter variables
than those chosen by Huse to get rid of an ambiguous and physically
doubtful procedure of summation over boundary conditions. Instead, we
consider physically more transparent linear combinations of the local
height operators for given boundary conditions. As a result we obtain a
set of form factors for the $\Phi_{1,3}$ perturbation of the minimal
conformal models, as well as reproduce the expressions for form factors
of the sine-Gordon model, obtained earlier by
Lukyanov~\cite{Lukyanov96}. We obtain form factors in the form of
multiple integrals. Note that some simple expressions for two-particle
form factors in the scaling RSOS models were obtained by
Delfino~\cite{Delfino:1999xh}.

Another result of the paper is a derivation of the vacuum expectation
values of some primary operators in the minimal models of conformal
field theory from those of exponential operators of the sine-Gordon
model. The first conjecture on the vacuum expectation values was made
in~\cite{LukZam96}. Later it was revised~\cite{Fateev:1997yg}, and some
vacuum-dependent factor was added. Though the derivation of this paper
is based on some assumptions, it gives evidence that, up to signs, the
correct one is the former conjecture of \cite{LukZam96}. The
vacuum-dependent factor must be nothing but $\pm1$.

The paper is a kind of reflection on different aspects of the problem.
So the structure of the paper is rather loose. In Sec.~2 the basic
reference information on the SOS and RSOS models, their spaces of
states, vacuums, excitations, local operators and form factors is given.
The scaling limits of these models are described in Sec.~3. The scaling
limit of local height probabilities is obtained, and the normalization
of the corresponding operators in the field theory is discussed. Scaling
of the operators in the restricted theory is considered in Sec.~4.
Sec.~5 is devoted to derivation of the representation of the form
factors. In the Appendix an amusing arithmetic theorem concerning
enumeration of primary operators in the minimal models is proven.

%%%%%%%%%%%%%%%%%%%%%%%%%%%%%%%%%%%%%%%%%%%%%%%%%%%%%%%%%%%%%%%%%%%%%%%%
\section{SOS and RSOS models}

Consider the SOS model on the square lattice~\cite{Baxter73}. A variable
$n_i\in\Z+\delta$ with some fixed complex number $\delta$ is associated
to each vertex $i$ of the lattice, while the Boltzmann weights are
associated to each face of the lattice. The variables $n_i$ must satisfy
the admissibility condition
$$
|n_i-n_j|=1
$$
for adjacent lattice vertices $i$ and~$j$. The Boltzmann weights
$\W(n_1,n_2;n_3,n_4|u)$ of a face with local variables $n_1,\ldots,n_4$
surrounding the face are given by
\AlignStar{
\W(n\pm1,n\pm2;n\pm1,n|u)
&=R_0(u),
\\
\W(n\pm1,n;n\pm1,n|u)
&=R_0(u){[n\pm u][1]\over [n][1-u]},
\\
\W(n\mp1,n;n\pm1,n|u)
&=R_0(u){[n\pm1][u]\over [n][1-u]}
}
with square brackets defined in terms of the Jacobi theta
function $\theta_1(u;\tau)$ of quasiperiods $1$ and $\tau$:
$$
[u]=\sqrt{\pi\over\epsilon r}\,\e^{{1\over4}\epsilon r}\,
\theta_1\!\left({u\over r};{\i\pi\over\epsilon r}\right).
$$
The function $R_0(u)$ is an arbitrary normalization factor. The
parameters $r$ and $\epsilon$ label families of commuting transfer
matrices, while the parameter $u$ is the spectral parameter enumerating
the commuting transfer matrices in each family. The region
$$
\epsilon>0,
\qquad
r\ge1,
\qquad
0<u<1
$$
is called the regime~III.

The space of states $\cH^{(0)}_m$ of the SOS model in the regime~III
with the boundary condition $m$ is spanned by the vectors $|P\rangle$,
labeled by paths $P=\{n_k\}^\infty_{k=-\infty}$, such that
$n_k\in\Z+\delta$ and the following conditions are satisfied:
\eq{
\Aligned{
\text{(i)}\quad
&\text{$|n_{k+1}-n_k|=1$ (admissibility condition);}
\\
\text{(ii)}\quad
&\text{$n_{2k}=n_\infty$, $n_{2k+1}=n_\infty+1$
or $n_{2k}=n_\infty+1$, $n_{2k+1}=n_\infty$ for
$|k|\gg1$ (boundary condition).}
}
\label{PathCond}
}
In fact, not every value of $n_\infty$ is admissible. The value of
$n_\infty$ must satisfy the condition $Nr<\Re n_\infty<(N+1)r-1$ for
some integer~$N$. The subscript $m$ in the notation $\cH^{(0)}_m$ is
defined as follows:
$$
m=n_\infty-N.
$$

The space of states $\cH^{(0)}_m$ admits two integrability preserving
restrictions for particular values of the parameters of the model.
First, for $\delta=0$ it can be restricted to the space
$\cH^{(1)}_m\subset\cH^{(0)}_m$ spanned by the paths $P=\{n_k\}$, such
that
\eq{
\forall k:\ n_k>0,
\qquad
m>0.
\label{Red1}
}
In addition, for rational $r$,
\eq{
r={q\over q-p},
\qquad
q>p>0,
\qquad
\text{$q$ and $p$ being coprimes},
\label{Red2r}
}
it admits the restriction to the space $\cH^{(2)}_m\subset\cH^{(1)}_m$
spanned by the paths subject to
\eq{
\forall k:\ 0<n_k<q,
\qquad
0<m<p.
\label{Red2}
}
The models defined on the spaces $\cH^{(1)}_m$ and $\cH^{(2)}_m$ are
called {\it restricted\/} solid-on-solid (RSOS) models. To distinguish
the two situations we shall call them \RSOS1 and \RSOS2 models
respectively. At the critical point the SOS model gives the free boson
field theory in the scaling limit, while the \RSOS1 and \RSOS2 models
give conformal models with the central charge of the Virasoro algebra
\eq{
c=1-{6\over r(r-1)}.
\label{CentralCharge}
}
The \RSOS2 models give the minimal conformal models $M(p,q)$ with a
finite number $(p-1)(q-1)/2$ of primary fields~\cite{PN}. Below, we
shall denote by $\cH_m$ any of the spaces $\cH^{(i)}_m$ ($i=0,1,2$),
when we do not want to specify a restriction.

The eigenvector of the transfer matrix in the space $\cH_m$ of the
eigenvalue of the largest absolute value will be called the vacuum
$|\text{vac}\rangle_m$. It is convenient to normalize the weights so
that this largest eigenvalue would be equal to one, which is achieved by
appropriate choice of $R_0(u)$. Other eigenvectors of the transfer
matrix can be described in terms of elementary excitations (particles)
$A_s(\theta)$ with the `rapidity' $\theta$ and the `internal state' $s$.
The vector $|A_{s_1}(\theta_1)\ldots A_{s_N}(\theta_N)\rangle_m$
corresponds to the eigenvalue factorizing into a product
$\prod^N_{j=1}\tau_{s_j}(\theta_j)$ of one-particle functions
$\tau_s(\theta)$. There are two types of excitations in the SOS model:
kinks and breathers. Since breathers can be considered as bound states
of two kinks, we shall only describe the kinks. The one-kink function
is
$$
\tau_{\rm kink}(\theta)
=\tau(\theta+\i\pi u),
\qquad
\tau(\theta)
={\theta_4\!\left({1\over4}+{\theta\over2\pi\i};
{\i\pi\over2\epsilon}\right)
\over\theta_4\!\left({1\over4}-{\theta\over2\pi\i};
{\i\pi\over2\epsilon}\right)}.
$$
An internal state of a kink is described by a pair of admissible vacuum
labels $m_1$, $m_2$, $|m_2-m_1|=1$, so that the multi-kink eigenvectors
can be written as $|A(\theta_1)^{m_1}_mA(\theta_2)^{m_2}_{m_1}\ldots
A(\theta_N)^m_{m_{N-1}}\rangle_m$.

We shall study the scaling limit $\epsilon\to0$. In this limit
$$
\tau(\theta)=1+\i M\sh\theta+O(M^2),
\qquad
\epsilon\to0,
\qquad
-{\pi^2\over2\epsilon}<\theta<{\pi^2\over2\epsilon},
$$
where
\eq{
M=4\e^{-\pi^2/2\epsilon}
\label{KinkMass}
}
is the kink mass. Because of this particular form of the mass we shall
often use the parameter $M/4$, which will be always written down
explicitly to avoid a confusion.

Consider any operator $O$ acting in the space $\cH_m$. Its form factors
are defined as
$$
F(O|\theta_1,\ldots,\theta_N)_{mm_1\ldots m_{N-1}}
={}_m\langle\text{vac}|O|A(\theta_1)^{m_1}_mA(\theta_2)^{m_2}_{m_1}\ldots
A(\theta_N)^m_{m_{N-1}}\rangle_m.
$$

Below we need some objects defined on a half-line and related to the
corner transfer matrix picture. Define the space $\cH^{(0)}_{mn}$ of
semi-infinite paths $|\{n_k\}^\infty_{k=0}\rangle$ subject to
(\ref{PathCond}) with $n_0=n$. The spaces $\cH^{(1)}_{mn}$ and
$\cH^{(2)}_{mn}$ are defined by the additional conditions (\ref{Red1})
and (\ref{Red2}) correspondingly. Again, $\cH_{mn}$ will denote any of
these spaces. There are some operators defined on these spaces
$$
H:\cH_{mn}\to\cH_{mn},
\qquad
\Psi^*(\theta)^{m\pm1}_m:\cH_{mn}\to\cH_{m\pm1,n}
$$
called corner Hamiltonian and type~II vertex operators.%
\footnote{Since we limit our consideration by the one-point local height
operators, we do not need the type~I vertex operators.}
The operator $\e^{-2\pi H}$ is proportional to the product of the four
corner transfer matrices. The operator $H$ possesses an equidistant
spectrum spaced by $2\epsilon/\pi$. Its eigenvalues are degenerate.
The multiplicities of this degeneration are given by the generating
functions
$$
\chi_{mn}(z)=\Tr_{\cH_{mn}}z^{\pi H/2\epsilon}
=z^{\Delta_{mn}}
\sum_{k=0}^\infty z^k\dim\cH_{mn}(k),
$$
where $\Delta_{mn}$ is the lowest eigenvalue of the operator $\pi
H/2\epsilon$ on the space $\cH_{mn}$, and $\cH_{mn}(k)$ is its
eigensubspace corresponding to the eigenvalue $\Delta_{mn}+k$. It turns
out~\cite{Andrews:1984af} that the minimal eigenvalues are equal to
\eq{
\Delta_{mn}={(rm-(r-1)n)^2-1\over4r(r-1)},
\label{Deltamn}
}
while the generating functions in the three cases are given by
\Align{
\chi^{(0)}_{mn}(z)
&={z^{\Delta_{mn}}\over\prod^\infty_{k=1}(1-z^k)},
\label{Chi0}
\\
\chi^{(1)}_{mn}(z)
&=\chi^{(0)}_{mn}(z)-\chi^{(0)}_{m,-n}(z),
\label{Chi1}
\\
\chi^{(2)}_{mn}(z)
&=\sum_{k\in\Z}(\chi^{(0)}_{m,n+2qk}(z)-\chi^{(0)}_{m,-n+2qk}(z)).
\label{Chi2}
}
Note, that for $m$ and $n$ being integers this formula gives highest
weights of the degenerate representations of the Virasoro algebra with
the central charge given by~(\ref{CentralCharge}). The characters
$\chi^{(1)}_{mn}(z)$ and $\chi^{(2)}_{mn}(z)$ coincide with the
characters of the respective irreducible representations of the Virasoro
algebra.

For future use we introduce the notation
\eq{
\alpha_+=\sqrt{2{r\over r-1}},
\qquad
\alpha_-=-\sqrt{2{r-1\over r}},
\qquad
\alpha_0={\alpha_++\alpha_-\over2}
={1\over\sqrt{2r(r-1)}}
\label{alpha_+-}
}
and
\eq{
\alpha_{mn}={1-m\over2}\alpha_++{1-n\over2}\alpha_-.
\label{alpha_mn}
}
With this notation we have
\eq{
c=1-12\alpha_0^2,
\qquad
\Delta_{mn}={(\alpha_{mn}-\alpha_0)^2-\alpha_0^2\over2}.
\label{c-Delta-alpha}
}

It turns out that for $m,n\in\Z_{>0}$ the space $\cH^{(0)}_{m,-n}$ can
be immersed into $\cH^{(0)}_{mn}$ as a subspace, so that
$\cH^{(1)}_{mn}\simeq\cH^{(0)}_{mn}/\cH^{(0)}_{m,-n}$. Similarly,
$\cH^{(2)}_{mn}\simeq\cH^{(0)}_{mn}/(\cH^{(0)}_{m,-n}
\cup\cH^{(0)}_{m,2q-n})$. Then the formulas (\ref{Chi1}) and
(\ref{Chi2}) are generalized to the identities
\Align{
\Tr_{\cH^{(1)}_{mn}}X
&=\Tr_{\cH^{(0)}_{mn}}X-\Tr_{\cH^{(0)}_{m,-n}}X,
\label{Tr1}
\\
\Tr_{\cH^{(2)}_{mn}}X
&=\sum_{k\in\Z}(\Tr_{\cH^{(0)}_{m,n+2qk}}X
-\Tr_{\cH^{(0)}_{m,-n+2qk}}X)
\label{Tr2}
}
for any operator $X$ on the space $\cH^{(i)}_{mn}$ realized as a factor
space.

The quantities
$$
\chi_{mn}=\chi_{mn}(x^4)=\Tr_{\cH_{mn}}\e^{-2\pi H},
\qquad
x=\e^{-\epsilon},
$$
are of physical importance: they determine the local height
probabilities. Namely, let $\Pi_{mn}$ be the projector in the space
$\cH_m$ onto the subspace spanned by the vectors $|\{n_k\}\rangle$ with
$n_0=n$. The local height probabilities are
$$
P_{mn}=\langle\Pi_{mn}\rangle={[n]\chi_{mn}\over\sum_n[n]\chi_{mn}}.
$$
The sum in the denominator is taken over all admissible values of $n$,
namely, $n\in\Z+\delta$ in the case of the SOS model and over
$n\in\Z_{>0}$ and $n=1,2,\ldots,q-1$ in the cases \RSOS1 and \RSOS2
correspondingly.

The corner Hamiltonian and vertex operators satisfy the relations
\Align{
[H,\Psi^*(\theta)^{m'}_m]
&=\i{d\over d\theta}\Psi^*(\theta)^{m'}_m,
\label{HPsiCommut}
\\
\Psi^*(\theta_1)^{m'}_s\Psi^*(\theta_2)^s_m
&=\sum_{s'}\S(s,m;s',m'|\theta_1-\theta_2)
\Psi^*(\theta_2)^{m'}_{s'}\Psi^*(\theta_1)^{s'}_m,
\label{PsiPsiCommut}
\\
\Psi^*(\theta')^{m'}_{m''}\Psi^*(\theta)^{m''}_m
&=-{\i[m'']'\over\theta'-\theta-\i\pi}\delta_{m'm}+O(1)
\text{ as $\theta'\to\theta+\i\pi$}.
\label{PsiNorm}
}
Here $[u]'=[u]|_{r\to r-1}$, and the kink $S$ matrix
\eq{
\S(m_1,m_2;m_3,m_4|\theta)
=-\left.\W(m_3,m_2;m_1,m_4|{\theta\over\i\pi})\right|_{r\to r-1}
\label{SOS_S_matrix}
}
is expressed in terms of the Boltzmann weights with the normalization
factor $R_0(u)$ chosen so that the largest eigenvalue is equal to one
for $0<u<1$, and with the shifted parameter~$r$. The products of type~II
vertex operators on the half line represent many-kink states in the
space of states $\cH_m$ on the full line.

The form factors of the local height operator $\Pi_{mn}$ are known to be
given by the trace function
\eq{
F(\Pi_{mn}|\theta_1,\ldots,\theta_N)_{mm_1\ldots m_{N-1}}
={[n]\over[m]'\chi}
\Tr_{\cH_{mn}}(e^{-2\pi H}
\Psi^*(\theta_N)^m_{m_{N-1}}\ldots\Psi^*(\theta_1)^{m_1}_m)
\label{TraceFunction}
}
with
\eq{
[m]'\chi=\sum_n[n]\chi_{mn}.
\label{Characters}
}
The value of $\chi$ is the same for all three cases of SOS, \RSOS1
and \RSOS2 models:
$$
\chi={2x^{-1/r(r-1)}\over\prod^\infty_{k=0}(1-x^{2+4k})}.
$$

%%%%%%%%%%%%%%%%%%%%%%%%%%%%%%%%%%%%%%%%%%%%%%%%%%%%%%%%%%%%%%%%%%%%%%%%
\section{Scaling limits in the unrestricted case: the problem of
identification and normalization of fields}

The unrestricted SOS model admits two different types of scaling limits:
\Align{
{\rm I}&:\ \epsilon\to0,
\qquad
\delta=\const,
\label{ScalingI}
\\
{\rm II}_\pm&:\ \epsilon\to0,
\qquad
\delta=\pm{\i\pi\over2 r\epsilon}.
\label{ScalingII}
}
Consider the first scaling limit (\ref{ScalingI}), which is consistent
with the restrictions for $\delta=0$. In this limit the $S$ matrix
(\ref{SOS_S_matrix}) tends to the following one
\eq{
\Aligned{
\S(m\pm1,m\pm2;m\pm1,m|\theta)
&=S_0(\theta),
\\
\S(m\pm1,m;m\pm1,m|\theta)
&=S_0(\theta){\sh{\i\pi m\pm\theta\over r-1}\sh{\i\pi\over r-1}
\over\sh{\i\pi m\over r-1}\sh{\i\pi-\theta\over r-1}},
\\
\S(m\mp1,m;m\pm1,m|\theta)
&=S_0(\theta){\sh{\i\pi(m\mp1)\over r-1}\sh{\theta\over r-1}
\over\sh{\i\pi m\over r-1}\sh{\i\pi-\theta\over r-1}},
}\label{SOS_S_matrix_lim}
}
where
\eq{
S_0(\theta)=-\exp\left(
2\i\int^\infty_0{dt\over t}\,
{\sh{\pi t\over2}\sh{\pi(r-2)t\over2}\sin\theta t
\over\sh\pi t\sh{\pi(r-1)t\over2}}
\right).
\label{S0_factor}
}
%This $S$ matrix is related to the $S$ matrix of the sine-Gordon model by
%means of the vertex--face correspondence, basically described
%in~\cite{Reshetikhin:1989qg}. The $S$ matrix (\ref{SOS_S_matrix_lim})
%provides a twisted description of the sine-Gordon model, which is
%convenient for restrictions. Note that here we use a basis of elementary
%excitations different from that used in~\cite{Reshetikhin:1989qg}, which
%leads to a different set of form factors. Of course, the physical
%quantities, like correlation functions, obtained from both sets of form
%factors must coincide.

Consider now one of the scaling limits (\ref{ScalingII}), e.~g., the
${\rm II}_-$ limit. Because of a nonzero value of $\delta$ this scaling
limit is inconsistent with the restrictions. The $S$ matrix
(\ref{SOS_S_matrix}) tends, up to a simple factor, to the kink $S$
matrix $S_{SG;\beta}(\theta)$ of the sine-Gordon model with the action
\eq{
{\cal A}=\int d^2x\,\left(
{(\partial_\mu\varphi)^2\over8\pi}
+\mu\cos\beta\varphi
\right),
\qquad
\beta=\sqrt{2{r-1\over r}}=-\alpha_-.
\label{SG_action}
}
Namely,
\eq{
\S(m_1,m_2;m_3,m_4|\theta)
\to\e^{\theta{\ve_1-\ve'_1\over2(r-1)}}
\e^{\i\pi{\ve_1-\ve'_2\over2(r-1)}}
S_{SG;\beta}(\theta)^{\ve'_1\ve'_2}_{\ve_1\ve_2}
\label{S_to_S_SG}
}
with $\ve_1=m_1-m_2$, $\ve_2=m_4-m_1$, $\ve'_1=m_4-m_3$,
$\ve'_2=m_3-m_2$. The parameter $\mu$ in the action is related to the
kink mass~\cite{Zamolodchikov:1995xk} as
\eq{
\mu=(M/4)^{2/r}\check\mu,
\qquad
\check\mu={2\over\pi}{\Gamma(1-1/r)\over\Gamma(1/r)}
\left(2\sqrt\pi\Gamma(r/2)\over\Gamma((r-1)/2)\right)^{2/r}.
\label{mu_M_relation}
}

The $S$ matrix (\ref{SOS_S_matrix_lim}), (\ref{S0_factor}) of the
scaling SOS model with finite $\delta$ is related to the $S$ matrix of
the sine-Gordon model $S_{SG;\beta}(\theta)$ by means of the vertex-face
correspondence, basically described in~\cite{Reshetikhin:1989qg}. From
the first glance, it seems that both describe the same model in
different bases of elementary excitation, but it is not quite correct.
To understand it better, let us recall the vertex-face correspondence
between the lattice SOS model and the eight-vertex
model~\cite{Baxter73}. The eight-vertex model corresponds to the
compactification of the sine-Gordon model, where
$\varphi\sim\varphi+4\pi/\beta$ so that the fields $\e^{\i
l\beta\varphi/2}$ with $l\in\Z$ are the only local neutral exponential
operators. The vacuum of this model is double degenerated while the
degeneracy of the SOS model is infinite. On the other hand, every
eigenvector of the transfer matrix of the SOS model is mapped onto an
eigenvector in the eight-vertex model with the same eigenvalue. It means
that the SOS model is an extension of the eight-vertex model, where each
vector of the eight-vertex model is infinitely repeated in the SOS
model. Hence, the scaling SOS model is an extension of the compactified
sine-Gordon model. More precisely, we have a one-parametric family
(parameterized by~$\delta\in\C/\Z$) of extensions of the compactified
sine-Gordon model. For $\delta=0$ the extension admits one or two
restrictions described above. In the limits $\delta\to\pm\infty$ the
extention coincides with the usual non-compact sine-Gordon model. The
most important fact, which will be used below, is that the vacuums
$|{\rm vac}\rangle_m$ for finite $\delta$ are not linear combinations of
the vacuums of the full sine-Gordon theory, since they are living,
strictly speaking, in another theory. In particular, the restricted
sine-Gordon model is the restriction just of the $\delta=0$ scaling SOS
model rather than that of the genuine sine-Gordon model.

Let us start our analysis from the unrestricted SOS model. Instead of
local height operators themselves it is more convenient to consider
their Fourier transforms:
$$
\Pi_m(a)=\sum_{n\in\Z+m}{\e^{\i\pi{an\over r}}
\over\sin{\pi n\over r}}\,\Pi_{mn}.
$$
The denominator $\sin{\pi n\over r}$ is introduced to get rid (in the
scaling limit) of the $n$ dependence related to the factor $[n]$ before
the trace in the form factor (\ref{TraceFunction}).

First, consider the vacuum expectation value of the operator
$\Pi_m(a)$:
$$
P_m(a)=\langle\Pi_m(a)\rangle
=\sum_{n\in\Z+m}{[n]\chi^{(0)}_{mn}\over[m]'\chi^{(0)}_m}
{\e^{\i\pi{an\over r}}\over\sin{\pi n\over r}}.
$$
Let us calculate $P_m(a)$ in the first scaling limit~(\ref{ScalingI}).
In this limit
$$
[u]\simeq\sqrt{\pi\over\epsilon r}\,\e^{-\pi^2/4\epsilon r}
\sin{\pi u\over r},
\qquad
\prod^\infty_{n=0}{1-x^{2+4n}\over1-x^{4+4n}}
\simeq2\sqrt{\epsilon\over\pi}.
$$
Substituting $n=m+k$ we obtain
$$
P_m(a)\simeq\sqrt{{\epsilon\over\pi}{r-1\over r}}\,
{\exp\left(\i\pi{am\over r}+{\pi^2\over4\epsilon r(r-1)}
-{\epsilon m^2\over r(r-1)}\right)\over\sin{\pi m\over r-1}}
\sum_{k\in\Z}\e^{-\epsilon{r-1\over r}k^2+2\epsilon{mk\over r}
+\i\pi{ak\over r}}
$$
The sum in the r.~h.~s.\ is a theta function, which admits the modular
transformation:
$$
\sum_{k\in\Z}\e^{\i\pi\tau k^2+2\pi\i ku}
=(-\i\tau)^{-1/2}\sum_{k\in\Z}\e^{-\i\pi(u+k)^2/\tau}.
$$
Here $u={a\over2r}-{\i\epsilon m\over\pi r}$,
$\tau={\i\epsilon\over\pi}{r-1\over r}$. In the limit $\tau\to0$ for
$-k-{1\over2}<\Re u<-k+{1\over2}$ the leading term in the r.~h.~s.\ is
$\e^{-\i\pi(u+k)^2/\tau}$. It means that for
\eq{
-r<\Re a+{2\epsilon\over\pi}\Im m<r
\label{genRegion}
}
the only term with $k=0$ survives. For real $m$ and $a$ the validity
region is
\eq{
-r<a<r.
\label{specRegion}
}

Finally, we have
\eq{
P_m(a)\simeq
{e^{\i\pi{am\over r-1}}\over\sin{\pi m\over r-1}}
\left(M\over4\right)^{a^2-1\over2r(r-1)},
\label{PmScaling}
}
where $M$ is the kink mass~(\ref{KinkMass}). The expectation
value of a local operator must be proportional to $M^{2\Delta}$ with
$\Delta$ being the conformal dimension of the operator. We conclude that
the conformal dimension of the operator $\Pi_m(a)$ is given by
$$
\Delta(a)={a^2-1\over4r(r-1)}.
$$
As the minimal dimension is equal to $-1/4r(r-1)$ we can think that this
model must be identified with a perturbation of the twisted free boson with
the energy-momentum tensor
$$
T(z)=-{1\over2}(\partial\varphi)^2+\i\alpha_0\partial^2\varphi.
$$
The central charge of the corresponding Virasoro algebra
$c=1-12\alpha_0^2$ coincides with~(\ref{CentralCharge}). The dimension
of the exponential operator $\e^{\i\alpha\varphi(x)}$ is equal to
\eq{
\Delta_\alpha={(\alpha-\alpha_0)^2-\alpha_0^2\over2}.
\label{Delta_alpha}
}
In particular, for $\alpha=\alpha_{mn}$ this expression gives the
conformal weights $\Delta_{mn}$.

Let us write the scaling limit of the operator $\Pi_m(a)$ as a value of
some local field $\Phi_\alpha(x)$ of the dimension $\Delta_\alpha$:
\eq{
\Pi_m(a)\simeq\Phi_\alpha(0),
\qquad
a={\alpha_0-\alpha\over\alpha_0}.
\label{aAlpha}
}
For small enough $|\alpha-\alpha_0|$ there are two such fields in the
theory: $\e^{\i\alpha\varphi(x)}$ and
$\e^{\i(2\alpha_0-\alpha)\varphi(x)}$. Hence, the most general form of
the operator $\Phi_\alpha(x)$ is
\eq{
\Phi_\alpha(x)=f_{\alpha,m}\e^{\i\alpha\varphi(x)}
+g_{\alpha,m}\e^{\i(2\alpha_0-\alpha)\varphi(x)}.
\label{Phi_genform}
}
Equivalently, we can write
\eq{
\e^{\i\alpha\varphi(x)}
={1\over2\i}
(A_{\alpha,m}\Phi_\alpha(x)+B_{\alpha,m}\Phi_{2\alpha_0-\alpha}(x)).
\label{exp_Phi}
}
The coefficients $f_{\alpha,m}$, $g_{\alpha,m}$ and $A_{\alpha,m}$,
$B_{\alpha,m}$ may depend on the
vacuum~$m$. From the fact that $\Phi_0(x)-\Phi_{2\alpha_0}(x)=2\i$ we get
\eq{
A_{0,m}=-B_{0,m}=1.
\label{AB_rel}
}
We shall fix the coefficients $A_{\alpha,m}$, $B_{\alpha,m}$ later.

Now we discuss the normalization of the operators $\Phi_\alpha(x)$. We
shall need this later to compare our results to the exact vacuum
expectation values~\cite{LukZam96,Fateev:1997yg}. The normalization of
local operators is usually fixed by the short range (ultraviolet)
asymptotics of their pair correlation functions. Note that the next
consideration concerns an {\it arbitrary\/} field of the form
(\ref{Phi_genform}), since it refers no other properties of the
particular field~(\ref{aAlpha}).

Consider first the operator product expansions of the exponential
operators. Let $\alpha_1$, $\alpha_2$ be arbitrary real numbers. Let
$\alpha_{12}(k,l)=\alpha_1+\alpha_2-2\alpha_{kl}$. Let $\Delta_1$,
$\Delta_2$, and $\Delta_{12}(k,l)$ be the corresponding conformal
dimensions according to the formula~(\ref{Delta_alpha}). From conformal
perturbation theory we know the following general form of the operator
product expansion of two exponential operators:
$$
\Aligned{
\e^{\i\alpha_1\varphi(x)}\e^{\i\alpha_2\varphi(y)}
&=\sum_{k,l\in\Z}
|x-y|^{2\Delta_{12}(k,l)-2\Delta_1-2\Delta_2}
D_{\alpha_1,\alpha_2}^{kl,m}(M|x-y|)
\e^{\i\alpha_{12}(k,l)\varphi(y)}
\\
&\quad
+\text{(contributions of descendants)}
}
$$
with some structure functions $D_{\alpha_1,\alpha_2}^{kl,m}(z)$. The
values $D_{\alpha_1,\alpha_2}^{kl,m}(0)$ are expressed in terms of the
multiple integrals over the Euclidean plane defined
in~\cite{Dotsenko:1984ad,Dotsenko:1988aspm}. Some of these coefficients
are very simple. In particular, we know that
$D_{\alpha_1,\alpha_2}^{11,m}(0)=1$. The other coefficients are rather
complicated. Fortunately, some information about these constants is
encoded in the vacuum expectation values of the exponential operators.

To fix normalization of the operators $\Phi_\alpha$ let us consider the
ultraviolet limit of the correlation functions
$\langle\e^{\i\alpha\varphi(x)}\e^{\i\alpha\varphi(y)}\rangle_m$,
$\langle\e^{\i\alpha\varphi(x)}
\e^{\i(2\alpha_0-\alpha)\varphi(y)}\rangle_m$. We are interesting in the
contributions proportional to $|x-y|^{-4\Delta_\alpha}$. For the latter
of the two functions we easily get
$$
\langle\e^{\i\alpha\varphi(x)}
\e^{\i(2\alpha_0-\alpha)\varphi(y)}\rangle_m
=|x-y|^{-4\Delta_\alpha}
\langle\e^{\i2\alpha_0\varphi(y)}\rangle_m+\ldots.
$$
The dots designate all other contributions into the short range
asymptotics. Indeed, there are two fields of the dimension 0, the unit
operator and $\e^{\i2\alpha_0\varphi}$, that can appear in the expansion
of the operator product in the l.~h.~s. But the coefficient at the unit
operator $D_{\alpha,2\alpha_0-\alpha}^{-1\>-1,m}(0)$ is known to be zero
for any value of~$\alpha$. Then the r.~h.~s.\ is determined by the
second operator.

The product
$\langle\e^{\i\alpha\varphi(x)}\e^{\i\alpha\varphi(y)}\rangle_m$ is more
complicated. Its part proportional to $|x-y|^{4\Delta_\alpha}$ depends
on the parameter $\alpha$ discontinuously. For generic $\alpha$ the
respective part vanishes. But for the discrete values
$\alpha=\alpha_{kl}$ ($k,l\in\Z$) we have
$$
\langle\e^{\i\alpha\varphi(x)}\e^{\i\alpha\varphi(y)}\rangle_m
=|x-y|^{-4\Delta_\alpha}D_{\alpha,\alpha}^{kl,m}(0)+\ldots.
$$
Here, on the contrary, the only nonzero contribution is that of the unit
operator, while the contribution of the operator
$\e^{\i2\alpha_0\varphi}$ vanishes:
$D_{\alpha,\alpha}^{k+2\>l+2,m}(0)=0$.

The value $D_{\alpha,\alpha}^{kl,m}(0)$ can be easily
evaluated using the `continuation' from the sinh-Gordon theory in the
spirit of~\cite{LukZam96,Fateev:1997yg}. In the sinh-Gordon theory,
where all $\alpha$s and $\beta$ are imaginary and the vacuum is unique,
we get
$$
\langle\e^{\i\alpha\varphi(x)}\e^{\i\alpha\varphi(y)}\rangle
=\langle\e^{\i\alpha\varphi(x)}\e^{\i(2\alpha_0-\alpha)\varphi(x)}\rangle
{\langle\e^{\i\alpha\varphi}\rangle
\over\langle\e^{\i(2\alpha_0-\alpha)\varphi}\rangle}+\ldots
=|x-y|^{-4\Delta_\alpha}{\langle\e^{\i\alpha\varphi}\rangle
\langle\e^{\i2\alpha_0\varphi}\rangle
\over\langle\e^{\i(2\alpha_0-\alpha)\varphi}\rangle}+\ldots.
$$
Though the first and second equalities taken separately are, of course,
wrong in the sine-Gordon theory, the final result must be valid for
the special fields:
\eq{
\langle\e^{\i\alpha\varphi(x)}\e^{\i\alpha\varphi(y)}\rangle_m
=|x-y|^{-4\Delta_\alpha}{\langle\e^{\i\alpha\varphi}\rangle_m
\langle\e^{\i2\alpha_0\varphi}\rangle_m
\over\langle\e^{\i(2\alpha_0-\alpha)\varphi}\rangle_m}+\ldots
\quad
\text{for $\alpha=\alpha_{kl}$.}
\label{alphaalphanorm}
}
Note that, despite of its form in terms of vacuum expectation values of
some massive theory, the coefficient in the r.~h.~s.\ is nothing but a
structure constant of a kind of conformal field theory, and the
`analytic continuation' is simply a roundabout way to express the result
of the integrations in a convenient form.

For the operators $\Phi_\alpha(x)$ of the form (\ref{Phi_genform}) we
obtain
$$
\Aligned{
\langle\Phi_\alpha(x)\Phi_\alpha(y)\rangle
&=|x-y|^{-4\Delta_\alpha}c_{\alpha,m}+\ldots,
\\
\langle\Phi_\alpha(x)\Phi_{2\alpha_0-\alpha(y)}\rangle
&=|x-y|^{-4\Delta_\alpha}c'_{\alpha,m}+\ldots,
}
$$
where
\eq{
c_{\alpha,m}
={\langle\e^{\i2\alpha_0\varphi}\rangle_m
\over\langle\e^{\i\alpha\varphi}\rangle_m
\langle\e^{\i(2\alpha_0-\alpha)\varphi}\rangle_m}
\times\Cases{
2f_{\alpha,m}g_{\alpha,m}\langle\e^{\i\alpha\varphi}\rangle_m
\langle\e^{\i(2\alpha_0-\alpha)\varphi}\rangle_m,
&\alpha\not\in\{\alpha_{kl}\},
\\
\left(
f_{\alpha,m}\langle\e^{\i\alpha\varphi}\rangle_m
+g_{\alpha,m}\langle\e^{\i(2\alpha_0-\alpha)\varphi}\rangle_m
\right)^2,
&\alpha\in\{\alpha_{kl}\},
}
\label{c-gen}
}
and
\eq{
c'_{\alpha,m}
={\langle\e^{\i2\alpha_0\varphi}\rangle_m
\over\langle\e^{\i\alpha\varphi}\rangle_m
\langle\e^{\i(2\alpha_0-\alpha)\varphi}\rangle_m}
\times\Cases{
(f_{\alpha,m}f_{2\alpha_0-\alpha,m}+g_{\alpha,m}g_{2\alpha_0-\alpha,m})
\langle\e^{\i\alpha\varphi}\rangle_m
\langle\e^{\i(2\alpha_0-\alpha)\varphi}\rangle_m,
&\alpha\not\in\{\alpha_{kl}\},
\\
\left(
f_{\alpha,m}\langle\e^{\i\alpha\varphi}\rangle_m
+g_{\alpha,m}\langle\e^{\i(2\alpha_0-\alpha)\varphi}\rangle_m
\right)\times
\\
\qquad
\left(
f_{2\alpha_0-\alpha,m}\langle\e^{\i(2\alpha_0-\alpha)\varphi}\rangle_m
+g_{2\alpha_0-\alpha,m}\langle\e^{\i\alpha\varphi}\rangle_m
\right),
&\alpha\in\{\alpha_{kl}\}.
}
\label{c'-gen}
}
For generic values of $\alpha$ these expressions are rather ugly, while
for special ones they take a nice form:
\eq{
\Aligned{
c_{\alpha,m}
&={\langle\e^{\i2\alpha_0\varphi}\rangle_m
\over\langle\e^{\i\alpha\varphi}\rangle_m
\langle\e^{\i(2\alpha_0-\alpha)\varphi}\rangle_m}
\langle\Phi_\alpha\rangle_m^2,
\\
c'_{\alpha,m}
&={\langle\e^{\i2\alpha_0\varphi}\rangle_m
\over\langle\e^{\i\alpha\varphi}\rangle_m
\langle\e^{\i(2\alpha_0-\alpha)\varphi}\rangle_m}
\langle\Phi_\alpha\rangle_m
\langle\Phi_{2\alpha_0-\alpha}\rangle_m
\qquad
\text{for $\alpha=\alpha_{kl}$}.
}\label{cc'spec}
}

Consider the scaling limit~${\rm II}_-$. For the expectation
values we have
\eq{
P_m(a)\simeq2\i\e^{\i\pi{(a-1)m\over r-1}}
\e^{-{\pi^2(a^2-1)\over4r(r-1)}}
=2\i\e^{-\i\pi{(a-1)s\over r-1}}
\left(M\over4\right)^{(a-1)^2/2r(r-1)}
=2\i\e^{2\pi\i{\alpha\over\beta}s}
\left(M\over4\right)^{\alpha^2},
\label{PmSecondLim}
}
where
$$
s=\delta-m
$$
is a finite integer, which labels the degenerate sine-Gordon vacuums.
Denote these vacuums as $|{\rm vac}\rangle_s$.

It means that the conformal dimension of the field corresponding to the
leading contribution to the operator $\Pi_m(a)$ in this limit
$$
\Delta'_\alpha={\alpha^2\over2}
$$
coincides with that of the exponential field $\e^{\i\alpha\varphi}$ in
the free field theory with the energy-momentum tensor
$T(z)=-{1\over2}(\partial\varphi)^2$.

In this limit, we can identify the operator $\Pi_m(a)$ with the
exponential operator $\e^{\i\alpha\varphi}$. The factor
$\e^{2\pi\i{\alpha\over\beta}s}$ in (\ref{PmSecondLim}) is interpreted
in terms of multiple vacuums in the sine-Gordon model. Namely, the
vacuum $|{\rm vac}\rangle_s$ is defined by the expectation value of the
field $\varphi(x)$:
\eq{
\langle\varphi(x)\rangle_s
\equiv{}_s\langle{\rm vac}|\varphi(x)|{\rm vac}\rangle_s
={2\pi s\over\beta},
\qquad
s\in\Z.
\label{varphiVEV}
}
Let us fix the normalization of the exponential operators by the
ultraviolet limit of their pair correlation functions:
\eq{
\langle\e^{\i\alpha\varphi(x)}\e^{-\i\alpha\varphi(y)}\rangle_s
\simeq|x-y|^{-2\alpha^2}
\quad
\text{as}
\quad
|x-y|\to0.
\label{ExpNormalization}
}
With this normalization condition, the vacuum expectation values of the
exponential operators are known to be~\cite{LukZam96}
\Align{
\langle\e^{\i\alpha\varphi(x)}\rangle_s
&=G_\alpha\e^{2\pi\i{\alpha\over\beta}s}
\equiv(M/4)^{\alpha^2}\vG_\alpha\e^{2\pi\i{\alpha\over\beta}s},
\text{ where}
\notag
\\
\vG_\alpha
&=\left(2\sqrt\pi{\Gamma(r/2)\over\Gamma((r-1)/2)}\right)^{\alpha^2}
\exp\int^\infty_0{dt\over t}\,
\left(
{\sh^2\alpha\beta t\over
2\sh{\beta^2t\over2}\ch{(2-\beta^2)t\over2}\sh t}
-\alpha^2\e^{-2t}
\right).
\label{ExpVEV}
}
The phase factor $\e^{2\pi\i{\alpha\over\beta}s}$ omitted
in~\cite{LukZam96} is an immediate consequence of~(\ref{varphiVEV}).
Therefore, in the limit ${\rm II}_-$ we have
\eq{
\Phi_\alpha(x)=2\i
\vG_\alpha^{-1}\e^{\i\alpha\varphi(x)}.
\label{Phi-Exp-II-}
}

Our next aim is to generalize this result to arbitrary values of
$\delta$. First, generalize the expression for the vacuum expectation
value of the exponential field~(\ref{ExpVEV}). We conjecture that
\eq{
\langle\e^{\i\alpha\varphi(x)}\rangle_m
=(M/4)^{2\Delta_\alpha}\vG_\alpha\e^{-2\pi\i{\alpha\over\beta}m}.
\label{ExpVEVgen}
}
All off-diagonal vacuum matrix elements of the exponential operators are
supposed to be zero. Note that the last conjecture implies the fact that
the vacuums $|{\rm vac}\rangle_m$ are not any linear combinations of the
vacuums $|{\rm vac}\rangle_s$, otherwise some off-diagonal vacuum matrix
elements would be inevitably non-zero.

Now we want to find such values of the coefficients $A_{\alpha,m}$,
$B_{\alpha,m}$ that the equations (\ref{PmScaling}), (\ref{AB_rel}) were
satisfied. Besides, we demand consistency with the relation
(\ref{Phi-Exp-II-}) and a similar relation in the limit ${\rm II}_+$.
The solution is
\eq{
A_{\alpha,m}=\vG_\alpha,
\qquad
B_{\alpha,m}=-\vG_\alpha\e^{-4\pi\i{\alpha\over\beta}m}.
\label{ABConjecture}
}
This means that
\Align{
\e^{\i\alpha\varphi(x)}
&={\vG_\alpha\over2\i}(\Phi_\alpha(x)
-\e^{-4\pi\i{\alpha\over\beta}m}\Phi_{2\alpha_0-\alpha}(x)),
\label{Exp-Phi}
\\
\Phi_\alpha(x)
&={\e^{4\pi\i{\alpha_0\over\beta}m}
\over\vG_\alpha\sin(4\pi{\alpha_0\over\beta}m)}
\bigl(
\e^{\i\alpha\varphi(x)}
+\vR(\alpha)\e^{-4\pi\i{\alpha\over\beta}m}
\e^{\i(2\alpha_0-\alpha)\varphi(x)}
\bigr).
\label{Phi-Exp}
}
The function
\eq{
\vR(\alpha)
={\vG_\alpha\over\vG_{2\alpha_0-\alpha}}
=\left(
2\sqrt\pi
\left(r\over r-1\right)^r
{\Gamma\left(r\over2\right)
\over\Gamma\left({r-1\over2}\right)}
\right)^{{4\alpha_0(\alpha-\alpha_0)}}
{\Gamma(1-\alpha_+(\alpha-\alpha_0))
\Gamma(1-\alpha_-(\alpha-\alpha_0))
\over
\Gamma(1+\alpha_+(\alpha-\alpha_0))
\Gamma(1+\alpha_-(\alpha-\alpha_0))},
\label{vR(alpha)}
}
is a part of the so called reflection
function~\cite{Zamolodchikov:1995aa}, which plays an important role in
the Liouville field theory:
\eq{
R(\alpha)
={G_\alpha\over G_{2\alpha_0-\alpha}}
=(M/4)^{4\alpha_0(\alpha-\alpha_0)}\vR(\alpha).
\label{ReflectionFactorization}
}

In the limit ${\rm II}_-$ the first term in (\ref{Phi-Exp}) prevails
while in the limit ${\rm II}_+$ it is suppressed in comparison to the
second one. We have
$$
\Phi_\alpha(x)=
\Cases{2\i\vG_\alpha^{-1}\e^{\i\alpha\varphi(x)}
&\text{for $\delta=-\i\pi/2r\epsilon$ (the limit ${\rm II}_-$)},
\\
-2\i\vG_{2\alpha_0-\alpha}^{-1}
\e^{4\pi\i{2\alpha_0-\alpha\over\beta}m}\e^{\i(2\alpha_0-\alpha)\varphi(x)}
&\text{for $\delta=+\i\pi/2r\epsilon$ (the limit ${\rm II}_+$)}.
}
$$
This conforms (\ref{Phi-Exp-II-}) in the limit ${\rm II}_-$, while in
the limit ${\rm II}_+$ the role of the field $\varphi(x)$ is played by
the combination
\eq{
\varphi'(x)=4\pi m/\beta+\varphi(x),
\label{varphi'}
}
which possesses finite vacuum expectation values in this limit.

Substituting the equation (\ref{Phi-Exp}) into Eqs.~(\ref{c-gen}),
(\ref{c'-gen}) we obtain the explicit form of the normalization factors:
\eq{
c_{\alpha,m}={\vR(\alpha)\over\vG_\alpha^2\vR(0)}
\times\Cases{
2\left(
\e^{2\pi\i{\alpha_0-\alpha\over\beta}m}
\over\sin(4\pi{\alpha_0\over\beta}m)
\right)^2,
&\alpha\not\in\{\alpha_{kl}\},
\\
\left(
\e^{2\pi\i{\alpha_0-\alpha\over\beta}m}
\over\sin(2\pi{\alpha_0\over\beta}m)
\right)^{2\strut},
&\alpha\in\{\alpha_{kl}\},
}
\label{c_alpha,m}
}
and
\eq{
c'_{\alpha,m}={\vR(\alpha)\over\vG_\alpha^2\vR(0)}
\times\Cases{
{2\cos(4\pi{\alpha_0\over\beta}m)
\over\sin^2(4\pi{\alpha_0\over\beta}m)},
&\alpha\not\in\{\alpha_{kl}\},
\\
{1\over\sin^2(2\pi{\alpha_0\over\beta}m)},
&\alpha\in\{\alpha_{kl}\},
}
\label{c'_alpha,m}
}

%%%%%%%%%%%%%%%%%%%%%%%%%%%%%%%%%%%%%%%%%%%%%%%%%%%%%%%%%%%%%%%%%%%%%%%%
\section{Scaling limit in the restricted case: invariant operators,
their normalization and vacuum expectation values}

Above we have considered unrestricted models. Now let us consider the
first restriction for $\delta=0$. It must contain the trace
$\Tr_{\cH^{(0)}_{mn}}-\Tr_{\cH^{(0)}_{m,-n}}$. Take the combination
\eq{
\tilde\Pi_m(a)={\Pi_m(a)-\Pi_m(-a)\over2\i}
=\sum_{n\in\Z+m}{\sin{\pi an\over r}\over\sin{\pi n\over r}}\,\Pi_{mn}.
\label{InvOp}
}
Evidently,
$$
\sum_{n\in\Z}[n]\Tr_{\cH^{(0)}_{mn}}\tilde\Pi_m(a)X
=\sum^\infty_{n=1}[n]{\sin{\pi an\over r}\over\sin{\pi n\over r}}
(\Tr_{\cH^{(0)}_{mn}}\Pi_{mn}X-\Tr_{\cH^{(0)}_{m,-n}}\Pi_{m,-n}X)
=\sum^\infty_{n=1}[n]\Tr_{\cH^{(1)}_{mn}}\tilde\Pi_m(a)X.
$$
It means that for $\delta=0$ the operators $\tilde\Pi_m(a)$ are
invariant with respect to the first restriction in Smirnov's sense.
Their form factors in the restricted theory \RSOS1 coincide with those
in the unrestricted theory.

In the scaling limit the vacuum expectation values of the operator
$\tilde\Pi_m(a)$ is
\eq{
\tilde P_m(a)\simeq{\sin{\pi am\over r-1}\over\sin{\pi m\over r-1}}
\left(M\over4\right)^{2\Delta(a)}
={\sin(2\pi{\alpha_0-\alpha\over\beta}m)
\over\sin(2\pi{\alpha_0\over\beta}m)}
\left(M\over4\right)^{2\Delta_\alpha}.
\label{tildePm}
}
Note that, since for $a\in(r-1)\Z$ the r.~h.~s.\ vanishes, the
approximation we use fails at these points.

In the scaling limit the operator $\tilde\Pi_m(a)$ corresponds to the
local field
\eq{
\tilde\Phi_\alpha(x)={\Phi_\alpha(x)-\Phi_{2\alpha_0-\alpha}(x)\over2\i}.
\label{tildePhi_alpha}
}
Let us find the operators invariant with respect to the second
restriction for rational values of $r$. To provide the subtractions that
correspond to the trace over the space ${\cH^{(2)}_{mn}}$ according to
the relation (\ref{Tr2}) let us find such values of $a$ that the
necessary subtractions take place directly in the trace:
$$
\sum_{n\in\Z}[n]\Tr_{\cH^{(0)}_{mn}}\tilde\Pi_m(a)X
=\sum_{n\in\Z}[n]\Tr_{\cH^{(2)}_{mn}}\tilde\Pi_m(a)X.
$$
Indeed, every term in (\ref{Tr2}) can be obtained from the term with
$\Tr_{\cH^{(0)}_{mn}}$ by a combination of two reflections, $n\to-n$ and
$n\to2q-n$, with simultaneous change of the sign at the trace. As we
have seen above, the first reflection property is satisfied for
$\tilde\Pi_m(a)$ for an arbitrary value of~$a$. The second reflection
property is satisfied, if
$$
\sin{\pi a(2q-n)\over r}=-\sin{\pi an\over r}.
$$
We arrive to the following quantization condition for the parameter $a$:
$$
{aq\over r}=\nu,
\qquad
\nu\in\Z,
\qquad
0<a<r,
\qquad
a\not\in(r-1)\Z.
$$
It means that
\eq{
a=a_\nu\equiv{\nu\over q-p},
\qquad
\nu=1,2,\ldots,q-1,
\qquad
\nu\not\in p\Z.
\label{aQuantization}
}
Let us represent the value $\nu$ in the standard form
\eq{
\nu=\pm(qk-pl),
\qquad
0<k<p,
\qquad
0<l<q.
\label{klEquation}
}
The numbers $k$ and $l$ can be restored from the value $\nu$ by a simple
algorithm (see Appendix). This means that the operator
$\tilde\Pi_m(a_\nu)$ describes the primary field $\phi_{kl}(x)$ of the
conformal dimension $\Delta_{kl}$ of the minimal model $M(p,q)$.
Evidently,
\eq{
a_\nu={\alpha_0-\alpha_{kl}\over\alpha_0}.
\label{a_nu_alpha_kl}
}
Then
\eq{
\tilde\Phi_{kl}(x)\equiv
\tilde\Phi_{\alpha_{kl}}(x)=N_{kl}\phi_{kl}(x)
\label{Phi_kl}
}
with some normalization factor $N_{kl}$. Of course, not every primary
field $\phi_{kl}(x)$ can be identified with one of the lattice operators
$\tilde\Pi(a_\nu)$ due to the limitation $0<\nu<q$. To extract the whole
set of primary fields $\phi_{kl}(x)$ in every minimal model $M(p,q)$ we
probably needed to study subleading terms in the scaling series
for~$P_m(a)$.

There are two interesting cases, where the relation between the
numbers $k$,~$l$ and the parameter~$\nu$ is simple. The first one is
the unitary series $M(p,p+1)$, where the local height operators describe
the fields on the diagonal of the Kac table:
$$
\tilde\Phi_{kk}(0)=\tilde\Pi_m(a_k),
\qquad
k,m=1,2,\ldots,p-1.
$$
The second case is the series $M(2,2N+1)$, which contains the Lee--Yang
model as a particular example ($N=2$). For this particular series the
whole set of primary fields is reproduced:
$$
\tilde\Phi_{1l}=\tilde\Pi_1(a_{2(N-l)+1}),
\qquad
l=1,2,\ldots,N.
$$

The normalization coefficients $N_{kl}$ in (\ref{Phi_kl}) can be
established using the relations (\ref{c_alpha,m}), (\ref{c'_alpha,m}).
Consider the product
$$
\langle\tilde\Phi_\alpha(x)\tilde\Phi_\alpha(y)\rangle
=|x-y|^{-4\Delta_\alpha}\tilde c_{\alpha,m}+\ldots.
$$
It is easy to check that
$$
\tilde c_{\alpha,m}={\vR(\alpha)\over\vG_\alpha^2\vR(0)}
\left(
\sin(2\pi{\alpha_0-\alpha\over\beta}m)
\over\sin(2\pi{\alpha_0\over\beta}m)
\right)^2
\quad\text{for $\alpha\in\{\alpha_{kl}\}$.}
$$
Let us normalize the operators $\phi_{kl}$ as follows:
\eq{
\langle\phi_{kl}(x)\phi_{kl}(y)\rangle
=|x-y|^{-4\Delta_{kl}}+\ldots.
\label{phi_klNorm}
}
Then we obtain the normalization constant
\eq{
N_{kl}={1\over\vG_{\alpha_{kl}}}
\left(\vR(\alpha_{kl})\over\vR(0)\right)^{1/2}
|S_{kl,m}|,
\qquad
S_{kl,m}={\sin(2\pi{|\alpha_0-\alpha_{kl}|\over\beta}m)
\over\sin(2\pi{\alpha_0\over\beta}m)}.
\label{N_kl}
}
This makes it possible to get the vacuum expectation values of the
operators $\phi_{kl}(x)$ using given vacuum expectation values for the
sine-Gordon model~(\ref{ExpVEV}). Let
\eq{
\langle\phi_{kl}(x)\rangle_m
=H_{kl,m}=(M/4)^{2\Delta_{kl}}\vH_{kl,m}.
\label{phiVEVdef}
}
By using the expectation value $\tilde P_m(a)$ from (\ref{tildePm}) we
obtain
\eq{
\vH_{kl,m}=Z_{kl,m}\vH_{\alpha_{kl}}.
\label{Hkl,m}
}
Here $Z_{kl,m}=\pm1$ is a sign factor and
\Align{
\vH_\alpha
&=\vG_\alpha\left(\vR(0)\over\vR(\alpha)\right)^{1/2}
=\left(
2\sqrt\pi{\Gamma((r+2)/2)\over\Gamma((r-1)/2)}
\right)^{2\Delta_\alpha}
Q\left(\alpha-\alpha_0\over\alpha_0\right),
\\
Q(\eta)
&=\exp\int^\infty_0{dt\over t}\,\left(
{\ch2t\sh(\eta-1)t\sh(\eta+1)t\over2\ch t\sh(r-1)t\sh rt}
-(\eta^2-1)\alpha_0^2\e^{-4t}
\right)
\label{Qeta}
}
is the normalization factor obtained in~\cite{LukZam96}. In fact, we are
unable to fix the sign factor $Z_{kl,m}$ since it depends on the
definition of signs in the structure constants. It is known to be
$(-1)^{(l-1)m}$ for the particular case of $k=1$ for the choice of
structure constants defined in~\cite{Dotsenko:1985hi}. This result was
obtained in different ways~\cite{Fateev:1997yg,Pugai02,Pugai04} and well
checked by numerical computations~\cite{Guida:1997fs,BBLPZ}. In the
general case it can be conjectured that
\eq{
Z_{kl,m}=\sign S_{kl,m},
\label{Zkl,m}
}
which is consistent with the abovementioned particular result. It is
important to note that for generic $k$, $l$ the result
(\ref{phiVEVdef}), (\ref{Hkl,m}) contradicts to the conjecture
of~\cite{Fateev:1997yg}, where the factor $S_{kl,m}$ itself appears in
the place of the sign factor $Z_{kl,m}$ in~(\ref{Hkl,m}).

%%%%%%%%%%%%%%%%%%%%%%%%%%%%%%%%%%%%%%%%%%%%%%%%%%%%%%%%%%%%%%%%%%%%%%%%
\section{Form factors of scaling operators}

To proceed with the form factors we need an explicit realization of the
corner Hamiltonian and the vertex operators~\cite{LukPug,MW}. Consider
a set of the oscillators $a_k$ ($k\in\Z$, $k\ne0$) and a pair of zero
mode operators $\P$ and $\Q$ with the commutation relations
$$
[\P,\Q]=-\i,
\qquad
[a_k,a_l]=k\,{\sh\epsilon k\sh\epsilon rk\over
\sh2\epsilon k\sh\epsilon(r-1)k}\,\delta_{k+l,0}.
$$
Define the vacuum states $|m,n\rangle$ ($m,n\in\Z+\delta$) by the
relations
$$
\P|m,n\rangle=P_{mn}|m,n\rangle,
\qquad
a_k|m,n\rangle=0\quad(k>0),
\qquad
P_{mn}={1\over2}(\alpha_+m+\alpha_-n).
$$
It makes it possible to define the Fock modules $\cF_{mn}$ as the span
of the vectors of the form
$$
a_{-k_1}\ldots a_{-k_N}|m,n\rangle
\quad(k_i>0).
$$
There are evidences that, at least for generic $\delta$, the
space $\cF_{mn}$ can be identified with $\cH_{mn}$, so that the trace in
(\ref{TraceFunction}) can be considered as the trace over the space
$\cF_{mn}$.

Introduce the operator
$$
H={2\epsilon\over\pi}
\left({\P^2-\alpha_0^2\over2}
+\sum^\infty_{k=1}{\sh2\epsilon k\sh\epsilon(r-1)k
\over\sh\epsilon k\sh\epsilon rk}a_{-k}a_k
\right).
$$
It is identified with the corner Hamiltonian $H$
in~(\ref{TraceFunction}). It is easy to check that
$\Tr_{\cF_{mn}}(z^{\pi H/2\epsilon})=\chi^{(0)}_{mn}(z)$.

Introduce the linear combinations
$$
\phi(z)={\alpha_+\over2}
(\Q-\i\P\log z)
+\sum_{k\ne0}{a_k\over\i k}\,z^{-k}.
$$
Introduce two operators
$$
V(\theta)=z^{r/4(r-1)}\lcolon\e^{\i\phi(z)}\rcolon,
\qquad
\bar V(\theta)=z^{r/(r-1)}
\lcolon\e^{-\i\phi(x^{-1}z)-\i\phi(xz)}\rcolon
$$
with
$$
z=e^{-2\epsilon\i\theta/\pi},
\qquad
x=e^{-\epsilon}.
$$
The normal ordering $\lcolon\ldots\rcolon$ places $a_k$ with $k>0$ to
the right of those with $k<0$ and $\P$ to the right of $\Q$. The
operator $\bar V(\theta)$ is used to introduce the screening operator
$$
X_m(\theta)=\int_C{d\gamma\over2\pi}\,\bar V(\gamma)F_m(\gamma-\theta),
\qquad
F_m(\gamma)
={[\gamma/\i\pi+1/2-m]'
\over[\gamma/\i\pi-1/2]'}.
$$
The contour $C$ goes over the period of the integrand from
$\theta-\pi^2/2\epsilon$ to $\theta+\pi^2/2\epsilon$ along the real
axis, but with an inflection: it goes above the point $\theta+\i\pi/2$
and below the point $\theta-\i\pi/2$.

The vertex operators are given by
$$
\Aligned{
\Psi^*(\theta)^{m+1}_m
&=V(\theta),
\\
\Psi^*(\theta)^{m-1}_m
&=\eta^{-1}V(\theta)X_m(\theta).
}
$$
Here the constant $\eta$ is chosen to satisfy the
condition~(\ref{PsiNorm}).

Our aim is to obtain the form factors of the operator $\Phi_\alpha(x)$,
which are given by the scaling limit of the form factors
$$
F_a(\theta_1,\ldots,\theta_N)_{mm_1\ldots m_{N-1}}
\equiv F(\Pi_m(a)|\theta_1,\ldots,\theta_N)_{mm_1\ldots m_{N-1}}.
$$
These form factors contain the sum over $n$ of the form factors
(\ref{TraceFunction}) with some coefficients. The $n$-dependence of the
trace in (\ref{TraceFunction}) is only contained in the zero mode
contribution to the integrand. It can be written as
$$
\e^{-2\epsilon P_{mn}^2}
\prod_{i\in I_+}\e^{-\i\epsilon\alpha_+P_{m_in}\theta_i/\pi}
\prod_{i\in I_-}\e^{\i\epsilon\alpha_+
(2P_{m_in}\gamma_i-P_{m_i-2,n}\theta_i)/\pi},
$$
where $I_\pm=\{i|m_{i+1}=m_i\pm1\}$, $m_0=m_N=m$. In the limit
$\epsilon\to0$ all $P_{m_in}$ can be substituted by $P_{mn}$ and it
reduces to
$$
\exp\left(-2\epsilon P_{mn}^2-\i\epsilon\alpha_+P_{mn}\Theta/\pi
\right),
$$
where
$$
\Theta=\sum^N_{j=1}\theta_j-2\sum^{N/2}_{s=1}\gamma_s.
$$
Hence, the form factors of $\Pi_m(a)$ are given by
\Multline{
F_a(\theta_1,\ldots,\theta_N)_{mm_1\ldots,m_{N-1}}
\simeq
\int\prod_{i\in I_-}
\left({d\gamma_i\over2\pi}\,F_{m_i}(\gamma_i-\theta_i)\right)
\left[
\Tr(\e^{-2\pi H}V_N\ldots V_1)
\over\eta^{N/2}\Tr(\e^{-2\pi H})
\right]_*
\\
\times
\sum_{n\in\Z+m}{[n]\chi^{(0)}_{mn}\over[m]'\chi^{(0)}_m}
{\e^{\i\pi{an\over r}}\over\sin{\pi n\over r}}
\e^{-\i\epsilon\alpha_+P_{mn}\Theta/\pi},
\notag
}
where
$$
V_i=\Cases{V(\theta_i),&i\in I_+\\
V(\theta_i)\bar V(\gamma_i),&i\in I_-}
$$
and $[\ldots]_*$ means that in all operators and traces we only take
into account the contribution of nonzero modes $a_k$. Let us simplify
this expression. Note, that the expression in the brackets has a
finite limit for $\epsilon\to0$. The functions $F_m(\gamma)$ also have a
finite limit:
$$
F_m(\gamma)\simeq{\sh{\gamma+\i\pi/2-\i\pi m\over r-1}
\over\sh{\gamma-\i\pi/2\over r-1}}.
$$
At last, the sum over $n$ reduces to
$$
\exp\left(-\i m{\epsilon\over\pi}{r\over r-1}\Theta\right)
P_m\left(a+{\epsilon r\over\pi^2}\Theta\right)
\simeq
{\e^{\i\pi{am\over r-1}}\over\sin{\pi m\over r-1}}
\left(M\over4\right)^{2\Delta(a)}
\e^{-a\Theta/2(r-1)}
$$
Finally, in the limit $\epsilon\to0$ we obtain in
\Multline{
F_a(\theta_1,\ldots,\theta_N)_{mm_1\ldots,m_{N-1}}
\\
=P_m(a)\int\prod_{i\in I_-}
\left({d\gamma_i\over2\pi}\,F_{m_i}(\gamma_i-\theta_i)\right)
\left[
\Tr(\e^{-2\pi H}V_N\ldots V_1)
\over\eta^{N/2}\Tr(\e^{-2\pi H})
\right]_*
\e^{-a\Theta/2(r-1)}.
\label{Flimiting}
}

It is convenient to rewrite the final answer in the form similar
to~(\ref{TraceFunction}). Introduce a continuous set of oscillators
$a(t)$ ($t\in\R$) with the commutation relations~\cite{JKM}
$$
[a(t),a(t')]=t{\sh{\pi t\over2}\sh{\pi rt\over2}
\over\sh\pi t\sh{\pi(r-1)t\over2}}\delta(t+t').
$$
Let $|0\rangle$ is the vacuum state defined as
$$
a(t)|0\rangle=0,
\qquad
t>0.
$$
It defines a Fock module $\cF$ generated by all $a(-t)$ ($t>0$).

Below we introduce the operators $\phi(\theta)$, $V(\theta)$, $\bar
V(\theta)$ in terms of $a(t)$ and the constant $\eta$, which have not to
be confused similar notations above in terms of $a_k$. Though they are
intended to describe the traces in (\ref{Flimiting}) in the scaling
limit, they differ from the lattice objects.

Let
$$
\phi(\theta)=\int^\infty_{-\infty}{dt\over t}\,a(t)\e^{i\theta t}.
$$
Traces of vertex operators contain integrals of the form
$\int^\infty_0dt\,f(t)$ with $f(t)$ possessing a pole at the point
$t=0$. According to the regularization procedure of~\cite{JKM} they must
be interpreted as
$$
\int^\infty_0dt\,f(t)=\int_{C_0}dt\,f(t){\log(-t)\over2\pi\i},
$$
where the contour $C_0$ goes from $+\infty$ above the cut $[0,+\infty)$,
then around $0$, and then to $+\infty$ below the cut.

Introduce the elementary exponentials
$$
\Aligned{
V(\theta)
&=\lcolon\e^{\i\phi(\theta)}\rcolon\,,
\\
\bar V(\theta)
&=\lcolon\e^{-\i\phi(\theta+\i\pi/2)-\i\phi(\theta-\i\pi/2)}
\rcolon\,,
}
$$
where the normal ordering puts $a(t)$ with $t>0$ to the right of
$a(-t)$. The screening operator reads
$$
X_m(a;\theta)
=\int_{C'}{d\gamma\over2\pi}\,
\bar V(\gamma)e^{a\gamma/(r-1)}
{\sh{\gamma-\theta+\i\pi/2-\i\pi m\over r-1}
\over
\sh{\gamma-\theta-\i\pi/2\over r-1}}.
$$
The contour $C'$ goes along the real axis above $\theta+\i\pi/2$ and
below $\theta-\i\pi/2$.

It amounts to the following prescription for the scaling type~II vertex
operators:
$$
\Aligned{
\Psi^*(a;\theta)^{m+1}_m
&=V(\theta)\e^{-a\theta/2(r-1)},
\cr
\Psi^*(a;\theta)^{m-1}_m
&=\eta^{-1}V(\theta)X_m(a;\theta)\e^{-a\theta/2(r-1)}.
}
$$
These operators $\e^{a\theta(m'-m)/2(r-1)}\Psi^*(a;\theta)^{m'}_m$
satisfy the relations (\ref{HPsiCommut}), (\ref{PsiPsiCommut}) with the
$S$ matrix (\ref{SOS_S_matrix_lim}). With the normalization constant
$$
\eta^{-1}=
{\e^{\alpha_+^2(C_E+\log\pi(r-1))}
\over\pi(r-1)^2}
{\Gamma\left(r\over r-1\right)
\over\Gamma\left(-{1\over r-1}\right)}
\exp\int^\infty_0{dt\over t}\,
{\sh{\pi t\over2}\sh{\pi rt\over2}
\over\sh\pi t\sh{\pi(r-1)t\over2}}\e^{-\pi t},
$$
where $C_E$ is the Euler constant, the vertex operators satisfy the
condition
$$
\Psi^*(a;\theta')^{m'}_{m''}\Psi^*(a;\theta)^{m''}_m
=-{\i\sin{\pi m'\over r-1}\over\theta'-\theta-\i\pi}\delta_{m'm}
+O(1)
\text{ as $\theta'\to\theta+\i\pi$.}
$$

Introduce the notation
$$
\llangle X\rrangle={\Tr_\cF\e^{-2\pi H}X\over\Tr_\cF\e^{-2\pi H}}.
$$
For the operator $X$ of the form $\prod V(\theta_i)\prod\bar
V(\gamma_j)$ this quantity is well defined for $z<1$ and easily
calculated by means of the Wick theorem.

The form factors are given by
\eq{
F_a(\theta_1,\ldots,\theta_N)_{mm_1\ldots,m_{N-1}}
=P_m(a)\llangle\Psi^*(a;\theta_N)^m_{m_{N-1}}
\ldots\Psi^*(a;\theta_1)^{m_1}_m\rrangle,
\label{FscalingPsi}
}
with $P_m(a)$ given by~(\ref{PmScaling}). By checking the cyclicity and
kinetic residue properties of this form factor we can make sure that the
operator $\Phi_\alpha(x)$ is local with respect to the excitation
creating operators. It means that all Euclidean correlation functions
that only contain the operators $\Phi_\alpha(x)$ at some points $x_i$
and the bosonic excitation creating operators are well-defined
single-valued functions of~$x_i$. Hence, the operator
$\tilde\Phi_\alpha(x)$, which possesses the form factors
\eq{
\tilde F_a(\theta_1,\ldots)_{mm_1\ldots}
={1\over2\i}\left(F_a(\theta_1,\ldots)_{mm_1\ldots}
-F_{-a}(\theta_1,\ldots)_{mm_1\ldots}\right),
\label{FtildePscalingPsi}
}
is also local with respect to excitation creating operators. On the
contrary, the operators $\e^{\i\alpha\varphi(x)}$ are essentially
nonlocal due to the $m$ dependence of the
coefficient~$B_{\alpha,m}$ (\ref{ABConjecture}). These operators are
only local in the ${\rm II}_-$ limit, while in the ${\rm II}_+$ limit
the operators $\e^{\i\alpha\varphi'(x)}$ with $\varphi'(x)$ defined in
(\ref{varphi'}) are local.

Let us discuss validity of the formulas (\ref{FscalingPsi}),
(\ref{FtildePscalingPsi}). The scaling limit was calculated in the
region $-r<a<r$. It coincides with the region of convergence of
integrals in $\gamma_i$. As analytic functions in $a$ the scaling form
factors (\ref{FscalingPsi}) and (\ref{FtildePscalingPsi}) can be
continued to the whole complex plane. One may conjecture that this
analytic continuation gives the form factors for any exponential field
$\e^{\i\alpha\varphi(x)}$ with the relation~(\ref{aAlpha}) in the
unrestricted theory. For the \RSOS1 and \RSOS2 models we may think that
the analytic continuation of the form factors for the operators
$\tilde\Pi_m(a)$ describes the form factors of all primary operators
$\phi_{kl}(x)$, which correspond to $\alpha=\alpha_{kl}$. To validate
this conjecture it would be necessary either to treat more accurately
the subleading contributions to the scaling form factors or to consider
some multipoint local height operators. We defer this task to future
work.

Let us return to the scaling limit ${\rm II}_-$. Redefine the $\Psi^*$
operators:
$$
\Psi^*_+(\alpha;\theta)
=\Psi^*(a;\theta)^{m+1}_m\times\e^{\theta/2(r-1)},
\qquad
\Psi^*_-(\alpha;\theta)
=\Psi^*(a;\theta)^{m-1}_m
\times2\i\e^{-\theta/2(r-1)}\e^{-\i\pi{m-1\over r-1}}.
$$
We get for the form factors
\Align{
F_a(\theta_1,\ldots,\theta_N)_{mm_1\ldots m_{N-1}}
&\simeq2\i\left(M\over4\right)^{\alpha^2}
\prod_{i\in I_+}\e^{\theta_i/2(r-1)}
\prod_{i\in I_-}\e^{-\theta_i/2(r-1)}\e^{-\i\pi{m_i-1\over r-1}}
\notag
\\
&\quad\times
\e^{2\pi\i{\alpha\over\beta}s}
\llangle
\Psi^*_{\ve_N}(\alpha;\theta_N)\ldots\Psi^*_{\ve_1}(\alpha;\theta_1)
\rrangle,
\qquad
\ve_i=m_i-m_{i-1}.
\label{FaSecondLim}
}
Explicitly, these operators are given by
$$
\Aligned{
\Psi^*_+(\alpha;\theta)
&=V(\theta)\e^{\alpha\theta/\beta},
\cr
\Psi^*_-(\alpha;\theta)
&=\eta^{-1}V(\theta)X(\alpha;\theta)\e^{\alpha\theta/\beta}
}
$$
with
$$
X(\alpha;\theta)
=\int_{C'}{d\gamma\over2\pi}\,
\bar V(\gamma)
{e^{-2\alpha\gamma/\beta}\over\sh{\gamma-\theta+\i\pi/2\over r-1}}.
$$
Physically, the second line of (\ref{FaSecondLim}) represents the form
factor of the field $\vG_\alpha^{-1}\e^{\i\alpha\varphi(x)}$. These form
factors are written for the $S$ matrix defined in~(\ref{S_to_S_SG}).
If we want to obtain the form factors for the pure sine-Gordon $S$
matrix $S_{SG;\beta}(\theta)$, we need to make a simple gauge
transformation. Finally, we reproduce Lukyanov's result~\cite{Lukyanov96}
for the exponential fields:
$$
F(\e^{\i\alpha\varphi}|\theta_1,\ldots,\theta_N)_{\ve_1\ldots\ve_N,s}
=\e^{2\pi\i{\alpha\over\beta}s}G_\alpha\llangle
\Psi^*_{\ve_N}(\alpha;\theta_N)\ldots\Psi^*_{\ve_1}(\alpha;\theta_1)
\rrangle.
$$
The factor $\e^{2\pi\i{\alpha\over\beta}s}$ makes the operator
$\e^{\i\alpha\varphi(x)}$ be formally local with respect to soliton
creating operators. This locality not physical. The definition of the
soliton creating operators in the non-compact theory in terms of the
field $\varphi$ inevitably provides cuts, at which all correlation
functions are discontinuous. If we try to match the correlation
functions at both banks of each cut, we shall obtain a continuous but
multivalued functions, which correspond to the known mutual
quasilocality of the neutral exponential operators and the soliton
creating operators.

%%%%%%%%%%%%%%%%%%%%%%%%%%%%%%%%%%%%%%%%%%%%%%%%%%%%%%%%%%%%%%%%%%%%%%%%

\section{Discussion}

We discussed the scaling limits in the SOS and RSOS models aiming to
clarify some problems in the corresponding integrable quantum field
theories. We saw that this approach makes it possible to obtain some
expressions for form factors and sheds some light on the relation
between the unrestricted and restricted sine-Gordon theories. Two
scaling limits were considered, one of which gave us the sine-Gordon
theory, while the second one admitted the restrictions to the perturbed
minimal models. The same operators in the lattice theory can be
identified in both limits with local operators in quantum field theory,
which allows one to relate their vacuum expectation values.

On the other hand the above consideration fetches out some fundamental
problems to be solved in future. First of all, the correct description
of the SOS vacuums (or, in other words, the quantum group invariant
vacuums) in the sine-Gordon theory is absent. This is closely related to
the fact that the vertex-face correspondence, which is studied in much
detail for local variables of the lattice models, is poorly studied on
the level of excitations.

The second problem is understanding the special fields in the sine-Gordon
theory. It is known from practice that many results concerning the {\it
special\/} fields in the minimal conformal field theory and sine-Gordon
theory can be obtained by formal `analytic continuation' of the
corresponding results for {\it generic\/} fields in Liouville and
sinh-Gordon theories. But there is no firm basis for this procedure.
Moreover, it is not clear why the operator product expansions in the
Liouville and sinh-Gordon theories give a continuous spectrum of fields,
while the corresponding expansions in the minimal CFTs and sine-Gordon
theory are discrete. Another aspect of this problem is how the $2/\beta$
shifts of the parameter of the exponential operators appear in the
sine-Gordon theory that only contains the $\cos\beta\varphi$
perturbation.

I hope that these notes will stimulate efforts to solve these problems.

%%%%%%%%%%%%%%%%%%%%%%%%%%%%%%%%%%%%%%%%%%%%%%%%%%%%%%%%%%%%%%%%%%%%%%%%

\section*{Acknowledgments}

I am grateful to A.~Belavin, A.~Litvinov, S.~Lukyanov, Y.~Pugai,
S.~Vergeles, A.~Zamolodchikov and Al.~Zamolodchikov for stimulating
discussions. The work was supported, in part, by INTAS under the grant
INTAS 03--51--3350, Russian Foundation for Basic Research under the
grants RFBR 05--01--01007 and RFBR--JSPS 05--01--02934YaF, and the
Program for Support of Leading Scientific Schools of Russian Government
under the grant \#2044.2003.2.

%%%%%%%%%%%%%%%%%%%%%%%%%%%%%%%%%%%%%%%%%%%%%%%%%%%%%%%%%%%%%%%%%%%%%%%%
\appendix

\section*{Appendix}
\setcounter{section}1
\setcounter{equation}0

In this Appendix we prove the

\begin{theorem}
Let $p$ and $q$ be coprimes such that $q>p>0$ and\/%
\footnote{The last condition for $\nu$ is excessive being the
consequence of the previous one.}
\eq{
\nu\in\Z,
\qquad
|\nu|\ne qa+pb\quad\forall a,b\in\Z_{\ge0},
\qquad
|\nu|<pq.
\label{nuConditions}
}
Then the equation
\eq{
qk-pl=\nu
\label{qkplnu}
}
has a unique solution in integers $k$, $l$ in the domain
\eq{
0<k<p,
\qquad
0<l<q.
\label{MainDomain}
}
\end{theorem}

In particular, the hypothesis of the Theorem is valid for the equation
(\ref{klEquation}) with the conditions~(\ref{aQuantization}).

Uniqueness of a solution is evident. Let us prove existence.

Let $\alpha=(p,q,k,l,\nu)\in\Z^5$. Consider transformations on $\Z^5$
preserving the form of the equation~(\ref{qkplnu}). Let $n$ be a
positive integer. Then we may rewrite the equation (\ref{qkplnu}) as
$$
p(l-nk)-(q-np)k=-\nu.
$$
It makes it possible to define the transformation
$$
T_n:\alpha\mapsto\alpha_1,\qquad
p_1=q-np,
\qquad
q_1=p,
\qquad
k_1=l-nk,
\qquad
l_1=k,
\qquad
\nu_1=-\nu.
$$
The numbers $p_1$, $q_1$ are again coprimes. This transformation is
invertible in integers, namely
$$
T_n^{-1}:\alpha_1\mapsto\alpha,
\qquad
p=q_1,
\qquad
q=p_1+nq_1,
\qquad
k=l_1,
\qquad
l=k_1+nl_1,
\qquad
\nu=-\nu_1.
$$
After the transformations $T_n$ we have
$$
q_1k_1-p_1l_1=\nu_1.
$$
Chose $n$ to satisfy the conditions
$$
0<q-pn<p
\quad\Leftrightarrow\quad
0<p_1<q_1.
$$
This condition determines the value of $n$ uniquely for $p>1$. Subject
to this condition we have $p_1<p$, $q_1<q$. Now chose $n_1$ so that
$0<q_1-p_1n_1<p_1$. Let $\alpha_2=T_{n_1}\alpha_1=T_{n_1}T_n\alpha$.
Iterate the procedure:
$$
\alpha_{t+1}=T_{n_t}\alpha_t,
\qquad
0<q_t-p_tn_t<p_t,
$$
till $p_N=1$. The last equation reads
\eq{
q_Nk_N-l_N=\nu_N\equiv(-1)^N\nu.
\label{qNkNlN}
}
The numbers $n$, $n_t$, $q_N$ can be easily defined by the simple
continued fraction
\eq{
{q\over p}=[n_0,n_1,n_2,\ldots,n_{N-1},n_N],
\qquad
n_0=0,
\qquad
n_N=q_N>1,
\label{qpFrac}
}
where we use the notation
$$
[a_0,a_1,a_2,a_3,\ldots]=a_0+{1\over\displaystyle a_1+
{\strut1\over\displaystyle a_2+
{\strut1\over\displaystyle a_3+\ldots}}}.
$$

It is easy to solve the equation (\ref{qNkNlN}) by fixing any value of
$k_N$ and taking $l_N=q_Nk_N-(-1)^N\nu$. Then we can
calculate $k$, $l$ using the inverse transformations as
$$
\alpha=T_n^{-1}T_{n_1}^{-1}\ldots T_{n_N}^{-1}\alpha_N.
$$
The question is if the answer belongs to the domain~(\ref{MainDomain}).
To solve this problem let us rewrite the conditions (\ref{MainDomain})
in terms of $\alpha_N$. First, let us write down $\alpha$ in terms of
$\alpha_t$:
\AlignStar{
k
&=A_tk_t+B_tl_t,
&p
&=A_tp_t+B_tq_t,
\\
l
&=C_tk_t+D_tl_t,
&q
&=C_tp_t+D_tq_t.
}
The coefficients $A_t,\ldots,D_t$ satisfy the recursion relations
\eq{
\Aligned{
A_{t+1}
&=B_t,
&C_{t+1}
&=D_t,
\\
B_{t+1}
&=A_t+n_tB_t,
&D_{t+1}
&=C_t+n_tD_t
}\label{ABCDIterate}
}
with the initial conditions $A_0=1$, $B_0=0$, $C_0=0$, $D_0=1$.

It is convenient to introduce the ratios
$$
\Aligned{
x_t
&={B_t\over A_t}=[n_{t-1},n_{t-2},\ldots,n_1]\ge1,
\\
y_t
&={D_t\over C_t}=[n_{t-1},n_{t-2},\ldots,n_1,n_0]\ge1.
}
$$
Hence,
\eq{
\Aligned{
x_t
&=n_{t-1}+x_{t-1}^{-1},
\\
y_t
&=n_{t-1}+y_{t-1}^{-1}.
}\label{xyIterate}
}
The conditions (\ref{MainDomain}) can be rewritten as
$$
\Aligned{
&0<x_N^{-1}k_N+l_N<x_N^{-1}+q_N,
\\
&0<y_N^{-1}k_N+l_N<y_N^{-1}+q_N.
}
$$
Let rewrite these conditions in terms of $k_N$ and $\nu_N=(-1)^N\nu$:
\eq{
\Aligned{
&(q_N+x_N^{-1})(k_N-1)<\nu_N<(q_N+x_N^{-1})k_N,
\\
&(q_N+y_N^{-1})(k_N-1)<\nu_N<(q_N+y_N^{-1})k_N.
}\label{nuNcond}
}
The regions defined by these two lines have nonempty intersection in
$\R$, if
$$
(q_N+x_N^{-1})(k_N-1)<(q_N+y_N^{-1})k_N,
\qquad
(q_N+y_N^{-1})(k_N-1)<(q_N+x_N^{-1})k_N.
$$
This can be considered as a weak compatibility condition for the
inequalities~(\ref{nuNcond}). It can be rewritten as
\eq{
-q_N-y_N^{-1}<(x_N^{-1}-y_N^{-1})k_N<q_N+x_N^{-1}.
\label{Compatibility}
}
Calculate the coefficient at $k_N$. We have
$$
x_t^{-1}-y_t^{-1}
=(y_t-x_t)x_t^{-1}y_t^{-1}
=-(x_{t-1}^{-1}-y_{t-1}^{-1})x_t^{-1}y_t^{-1},
\qquad
x_1^{-1}=0.
$$
Therefore,
$$
x_N^{-1}-y_N^{-1}
=(-1)^N\prod^N_{t=2}x_t^{-1}\prod^N_{t=1}y_t^{-1}
=(-1)^NB_N^{-1}D_N^{-1}.
$$
It is easy to check that for both even and odd $N$ the compatibility
condition (\ref{Compatibility}) in combination with the strictest of the
conditions of (\ref{nuNcond}) provides the condition
\eq{
|\nu|=|\nu_N|<{(q_N+x_N^{-1})(q_N+y_N^{-1})\over|x_N^{-1}-y_N^{-1}|}
=(q_N+x_N^{-1})(q_N+y_N^{-1})B_ND_N.
\label{AbsNu1}
}
It is evident that
\eq{
(q_N+x_N^{-1})B_N=p,
\qquad
(q_N+y_N^{-1})D_N=q.
\label{pqIdent}
}
Substituting (\ref{pqIdent}) to (\ref{AbsNu1}) we obtain the last of the
conditions~(\ref{nuConditions}):
$$
|\nu|<pq.
$$

Consider now the conditions (\ref{nuNcond}) more accurately and find the
compatibility conditions in integers. We want to find the values of
$\nu_N=\xi_{N+1}$ for which these inequalities are NOT satisfied. The
respective value of $k_N$ will be denoted as $\xi_N$ (the reason for
this notation will be clear later). Let us take, for definiteness, the
case $\nu_N\ge0$. Consider first the case $N\in2\Z$. In this case we can
take
\eq{
(q_N+y_N^{-1})\xi_N\le\xi_{N+1}\le(q_N+x_N^{-1})\xi_N.
\label{ineqN}
}
With the definition
$$
x_{N+1}=q_N+x_N^{-1},
\qquad
y_{N+1}=q_N+y_N^{-1}
$$
we can write it also as
$$
y_{N+1}\xi_N\le\xi_{N+1}\le x_{N+1}\xi_N.
$$
We have to solve this equation for $\xi_{N+1}$, $\xi_N$ in nonnegative
integers. Let us take
$$
\xi_{N+1}=q_N\xi_N+\xi_{N-1}.
$$
Then using (\ref{xyIterate}) the inequality (\ref{ineqN}) is rewritten
as
$$
(n_{N-1}+x_{N-1}^{-1})\xi_{N-1}\le\xi_N
\le(n_{N-1}+y_{N-1}^{-1})\xi_{N-1}
$$
or
$$
x_N\xi_{N-1}\le\xi_N\le y_N\xi_{N-1}.
$$
Iterating this procedure we get
\eq{
\xi_t=n_{t-1}\xi_{t-1}+\xi_{t-2}
\label{xiIterate}
}
with the conditions
$$
\Aligned{
y_t\xi_{t-1}\le\xi_t\le x_t\xi_{t-1}
&\quad\text{for $t\in2\Z+1$},
\\
x_t\xi_{t-1}\le\xi_t\le y_t\xi_{t-1}
&\quad\text{for $t\in2\Z$}.
}
$$
Note that for all values of $t$ these conditions are equivalent. Hence,
starting from $N=2\Z+1$ we can derive the same inequalities. From now on
we do not restrict the consideration to the case of even~$N$.

Taking into account that $x_0=0$, $y_0=+\infty$ we arrive to the
conditions
\eq{
\xi_{-1},\xi_0\ge0.
\label{xiConditions}
}
Express all $\xi_t$ in terms of $\xi_0$, $\xi_{-1}$. It is easy to prove
that
\eq{
\xi_t=B_t\xi_{-1}+D_t\xi_0.
\label{xiExpression}
}
Indeed, $\xi_1=\xi_{-1}+n\xi_0$ according to (\ref{xiIterate}), which
conforms~(\ref{xiExpression}). Besides, from (\ref{xiIterate}) we get
the recursion relation
$$
B_{t+1}=n_tB_t+B_{t-1},
\qquad
D_{t+1}=n_tD_t+D_{t-1},
$$
which conforms (\ref{ABCDIterate}).

Using (\ref{qpFrac}) and (\ref{pqIdent}) we get
$$
B_{N+1}=A_N+q_NB_N=(q_N+x_N^{-1})B_N=p,
\qquad
D_{N+1}=C_N+q_ND_N=(q_N+y_N^{-1})D_N=q.
$$
Finally we get that the values
$$
\xi_{N+1}=p\xi_{-1}+q\xi_0
\qquad
(\xi_{-1},\xi_0\ge0)
$$
are the only forbidden values of $|\nu_N|=|\nu|$, which proves the
theorem.

The constructive way to find $k$ and $l$ can be extracted from this
proof and formulated as

\begin{theorem}
Under the hypothesis of the Theorem~1 define the numbers
$n_0,\ldots,n_N$ by the simple continuous fraction
$$
{q\over p}=[n_0,n_1,\ldots,n_N],
\qquad
n_N>1.
$$
Consider any solution $k_N$, $l_N$ to the equation
$$
n_Nk_N-l_N=\nu_N\equiv(-1)^N\nu.
$$
Consider the recurrent relations
$$
k_{t-1}=l_t,
\qquad
l_{t-1}=k_t+n_{t-1}l_t.
$$
Let $k'=k_0$, $l'=l_0$. Then there exists such $s\in\Z$ that $k=k'-sp$,
$l=l'-sq$ is the solution to the equation~(\ref{qkplnu}) satisfying the
condition~(\ref{MainDomain}).
\end{theorem}

The proof is simple. Consider the transformation
$$
S:\alpha\mapsto\alpha',
\qquad
p'=p,
\qquad
q'=q,
\qquad
k'=k-p,
\qquad
l'=l-q,
\qquad
\nu'=\nu.
$$
It is easy to check that it commutes with $T_n$:
$$
T_nS=ST_n.
$$
We have proved that there exists a solution $k_N$, $l_N$ to the
equation~(\ref{qNkNlN}) corresponding to the solution $k=k_0$, $l=l_0$
to the equation~(\ref{qkplnu}) such that $0<k<p$, $0<l<q$. Evidently,
for any given solution $k'_N$, $l'_N$ to the equation~(\ref{qNkNlN})
there exists $s\in\Z$ such that
$(1,q_N,k_N,l_N,\nu_N)=S^s(1,q_N,k'_N,l'_N,\nu_N)$. Then
$(p,q,k,l,\nu)=S^s(p,q,k',l',\nu)$.

\end{document}